\documentclass[pra,a4paper,preprintnumbers,floatfix,aps,footinbib,showpacs,superscriptaddress,onecolumn,nofootinbib]{revtex4}

\usepackage{latexsym}
\usepackage{epsfig}
\usepackage{amssymb}
\usepackage{amsmath}

\newcommand{\bea}{\begin{eqnarray}}
\newcommand{\eea}{\end{eqnarray}}
\newcommand{\be}{\begin{equation}}
\newcommand{\ee}{\end{equation}}

\newcommand{\dd}{{\rm d}}
\newcommand{\Oo}{{\cal O}}

\newcommand\mpl{M_{\rm Pl}}
\newcommand{\GeV}{\,\mathrm{GeV}}
\newcommand{\meV}{\,\mathrm{meV}}

\newcommand{\Eqref}[1]{Eq.~\eqref{#1}}

\begin{document}

\title{Hidden in the Light: Magnetically Induced Afterglow from
  Trapped Chameleon Fields} 
\author{Holger Gies}
\affiliation{Institut f\"ur Theoretische Physik,
Universit\"at Heidelberg,  D-69120 Heidelberg, Germany}
\author{David F. Mota}
\affiliation{Institut f\"ur Theoretische Physik,
Universit\"at Heidelberg,   D-69120 Heidelberg, Germany}
\author{Douglas J. Shaw}
\affiliation{Astronomy Unit, School of Mathematical Sciences, Queen Mary, University of London, Mile End Road, London E1 4NS, United Kingdom}
\affiliation{DAMTP, Centre for Mathematical Sciences, University of Cambridge, Cambridge CB2 0WA, United Kingdom}

\begin{abstract}
  We propose an \emph{afterglow} phenomenon as a unique trace of
  chameleon fields in optical experiments. The vacuum interaction of a
  laser pulse with a magnetic field can lead to a production and
  subsequent trapping of chameleons in the vacuum chamber, owing to
  their mass dependence on the ambient matter density. Magnetically
  induced re-conversion of the trapped chameleons into photons creates
  an afterglow over macroscopic timescales that can conveniently be
  searched for by current optical experiments. We show that the
  chameleon parameter range accessible to available laboratory
  technology is comparable to scales familiar from astrophysical
  stellar energy loss arguments. We analyze quantitatively the
  afterglow properties for various experimental scenarios and discuss
  the role of potential background and systematic effects. We conclude
  that afterglow searches represent an ideal tool to aim at the production and
  detection of cosmologically relevant scalar fields in the laboratory.
\end{abstract}


\maketitle

\section{Introduction}

Light scalar fields populate theories of both cosmology and physics beyond the
Standard Model.  In generic models, these fields
can couple to matter and hence potentially mediate a new (or `fifth')
force between bodies.  However, no such new force has been detected
\cite{cwill}. Any force associated with such light scalar fields must
therefore be considerably weaker than gravity over the scales, and
under the conditions, that have been experimentally probed.  The fields must
either interact with matter far more weakly than gravity does, or be
sufficiently massive so as to have remained hidden thus far.  There is however an assumption when
deriving the above conclusions: the mass, $m$, of the scalar field is
taken to be a constant.  It has recently been shown that the most
stringent experimental limits on the properties of light scalar fields
can be exponentially relaxed if the scalar field theory in question
possesses a \emph{chameleon mechanism} \cite{chamKA,chamstrong}.  The
chameleon mechanism provides a way to suppress the forces mediated by
these scalar fields via non-linear field self-interactions.  A direct
result of these self-interactions is that the mass of the field is no
longer fixed but depends on, amongst other things, the ambient density
of matter. The properties of these scalar fields therefore change
depending on the environment; it is for this reason that such fields
have been dubbed \emph{chameleon fields}.\footnote{This is actually a
misnomer: despite popular belief, chameleons (the lizards) cannot
change color to their surroundings. Instead, changing color is an
expression of their physical and physiological condition
\cite{wiki:chameleon}.}. Chameleon fields could potentially also be
responsible for the observed late-time acceleration of the Universe
\cite{chamcos,chamstruc}. In the longer term, it has been shown that
future experimental measurements of the Casimir force will be able to
detect or rule out many, if not all, of the most interesting chameleon
field models for dark energy \cite{chamstrong,chamcas}. 

It was recently shown that some strongly-coupled (i.e., compared to
gravity) chameleon fields would alter light propagation through the
vacuum in the presence of a magnetic field in a polarization-dependent
manner \cite{chamPVLAS,chamPVLASlong}; the resultant birefringence and
dichroism could be detected in laboratory searches such as the
polarization experiments PVLAS
\cite{Zavattini:2005tm,Zavattini:2007ee}, Q\&A \cite{Chen:2006cd}, and
BMV \cite{Robilliard:2007bq} that are sensitive to new hypothetical
particles with a light mass and a weak coupling to photons. Popular
candidates for these searches are the axion \cite{Peccei:1977hh}, or
more generally an axion-like particle (ALP), minicharged particles
(MCPs) \cite{Okun:1982xi,Holdom:1985ag}, or paraphotons
\cite{Okun:1982xi}. These particle candidates may be viewed as
low-energy effective degrees of freedom of more involved microscopic
theories. In this sense, chameleons could be classified as ALPs as far
as optical experiments are concerned, but give rise to specific
optical signatures to be discussed in detail in this work.

In fact, a variety of further experiments such as ALPS
\cite{Ehret:2007cm}, LIPPS \cite{Afanasev:2006cv}, OSQAR \cite{OSQAR},
and GammeV \cite{GammeV} have been proposed and are currently built up
or are even already taking data. They look for anomalous optical
signatures from light propagation in a modified quantum vacuum. This
rapidly evolving field has been triggered by the fact that optical
experiments provide a rather unique laboratory tool, because photons
can be manipulated and detected with a great precision. Since
laboratory set-ups aim at both production and detection of new
particles under fully controlled experimental conditions, they are
complementary to astrophysical considerations such as those based on
stellar energy loss. The latter can indeed give rather strong bounds,
for instance, on axion models \cite{Raffelt:2006cw}. But for models
with particle candidates which exhibit very different properties in
the stellar plasma as compared with the laboratory environment, purely
laboratory-based experiments are indispensable
\cite{Masso:2005ym,chamstrong}. The chameleon is exactly of this type,
therefore being an ideal particle candidate for high-precision
laboratory-based searches.

This work is devoted to an investigation of possible optical
signatures which can specifically be attributed to chameleon fields,
thereby representing a smoking gun for this particle candidate. The
standard optical signatures in polarization experiments are induced
ellipticity and rotation for a propagating laser beam interacting with
a strong magnetic field \cite{Dittrich:2000zu}; these exist for a
chameleon \cite{chamPVLAS,chamPVLASlong}, but occur similarly for ALPs
\cite{Maiani:1986md}, MCPs \cite{Gies:2006ca} or models involving also
paraphotons \cite{Ahlers:2007rd}. In the case of a positive signal,
the various scenarios can be distinguished from each other by
analyzing the signal dependence on the experimental parameters such as
magnetic field strength, laser frequency, length of the interaction
region \cite{Ahlers:2006iz}. ALP and paraphoton models can
specifically be tested by light-shining-through-walls experiments
\cite{Sikivie:1983ip,Ahlers:2007rd}; MCPs can leave a decisive trace
in MCP production experiments in the form of a dark current
\cite{Gies:2006hv}.

In this work, we propose an \emph{afterglow} phenomenon as a unique chameleon
trace in an optical experiment.\footnote{A similar proposal can be
  found in \cite{DESYgroup}.} The existence of this afterglow
is directly linked with the environment dependence of the chameleon
mass parameter. In particular, the mass dependence on the ambient
matter density causes the trapping of chameleons inside the vacuum
chamber where they have been produced, e.g., by a laser pulse
interacting with a strong magnetic field. As detailed below, the
trapping can be so efficient that the re-conversion of chameleons into
photons in a magnetic field causes an afterglow over macroscopic time
scales. Most importantly, our afterglow estimates clearly indicate
that the new-physics parameter range accessible to current
technology is substantial and can become comparable to scales familiar from
astrophysical considerations. Afterglow
searches therefore represent a tool to probe physics halfway up to the
Planck scale.

This paper is organized as follows: In Sect.~\ref{sec:chammodel}, we review aspects of
the chameleon model which are relevant to the optical phenomena
discussed in this work. In Sect.~\ref{sec:chamjar}, we solve the equations of
motion for the coupled photon-chameleon system, paying particular
attention to the boundary conditions which give rise to the afterglow
phenomenon.  Signatures of the afterglow are exemplified in Sect.~\ref{sec:sigaglow}.
The chameleonic afterglow is compared with other background sources
and systematic effects in Sect.~\ref{sec:back}. Our conclusions are given in
Sect.~\ref{sec:con}. In the appendix, we discuss another option for
afterglow detection based on chameleon resonances in the vacuum
chamber.

\section{Chameleon Theories}\label{sec:chammodel}

As was mentioned above, chameleon theories are essentially scalar field theories with a self-interaction potential and a coupling to matter; they are specified by the action
\begin{eqnarray}
S&=&\int d^4x \sqrt{-g}\left(\frac{1}{2\kappa_4^2}R-
g^{\mu\nu}\partial_\mu\phi \partial_\nu \phi -V(\phi)\right)
\nonumber\\ 
&&+ S_m( e^{\phi/M_{i}}
g_{\mu\nu},\psi_{m}) -\frac{e^{\phi/M}}{4} F_{\mu\nu}F^{\mu\nu}
\label{action}, 
\end{eqnarray}
where $\phi$ is the chameleon field with a self-interaction
potential  $V(\phi)$. $S_{m}$ denotes the matter action and
$\psi_{m}$ are the matter fields, and we have also explicitly listed the
coupling to photons.  

The strength of the interaction between $\phi$ and the matter fields
is determined by the one or more mass scales $M_{i}$.  In general, we
expect different particle species to couple with different strengths
to the chameleon field, i.e., a different $M_{i}$ for each $\psi_{m}$.
Such a differential coupling would lead to violations of the weak
equivalence principle (WEP hereafter). It has been
shown that $V(\phi)$ can be chosen so that any violations of WEP are
too small to have been detected thus far \cite{chamstrong}.
Even though the $M_{i}$ are generally different for different species,
if $M_{i} \neq 0$, we generally expect $M_{i} \sim \Oo(M)$, with $M$
being some mass scale associated with the theory. Provided this is the case, the differential nature of the coupling will have very little effect on our predictions for the experiments considered here.  In this paper,  we therefore simplify the analysis by assuming a universal coupling $M_{i} = M$.

Note that if the matter fields are non-relativistic. The
scalar field, $\phi$, obeys:
\begin{equation}
\square \phi = V^{\prime}(\phi) + \frac{e^{\phi/M}\rho}{M}, \label{chameqn}
\end{equation}
where $\rho$ is the background density of matter.  The coupling to
matter implies that particle masses in the Einstein frame depend on
the value of $\phi$
\begin{equation}
m(\phi)= e^{\phi/M} m_0,
\end{equation}
where $m_0 = {\rm const}$ is the bare mass. 

We parametrize the strength of the chameleon to matter coupling by $\beta$ where
\begin{equation}
\beta= \frac{\mpl}{M},
\end{equation}
and $\mpl = 1/\sqrt{8\pi G} \approx 2.4 \times 10^{18}\GeV$. On
microscopic scales (and over sufficiently short distances), the
chameleon force between two particles is then $2\beta^2$ times the
strength of their mutual gravitational attraction.

If the mass, $m_{\phi} \equiv \sqrt{V^{\prime \prime}(\phi)}$, of
$\phi$ is a constant then one must either require that $m_{\phi}
\gtrsim 1\meV$ or $\beta \ll 1$ for such a theory not to have been
already ruled out by experimental tests of gravity
\cite{cwill}. If, however, the mass of the scalar field grows with the
background density of matter, then a much wider range of scenarios
have been shown to be possible \cite{chamKA,chamstrong,chamcos}.  In
high-density regions, $m_{\phi}$ can then be large enough so as to
satisfy the constraints coming from tests of gravity. At the same
time, the mass of the field can be small enough in low density regions
to produce detectable and potentially important alterations to
standard physical laws.  Assuming $\dd \ln m(\phi) / \dd \phi
\geq 0$ as it is above, a scalar field theory possesses a chameleon
mechanism if, for some range of $\phi$, the self-interaction
potential, $V(\phi)$, has the following properties:
\begin{equation}
V^{\prime}(\phi) < 0, \quad V^{\prime \prime} > 0,\quad V^{\prime \prime \prime}(\phi) < 0, \label{chamcond}
\end{equation}
where $V^{\prime} = \dd V / \dd\phi$.  The evolution of the chameleon
field in the presence of ambient matter with density $\rho_{\rm
matter}$ is then determined by the effective potential:
\begin{equation}
V_{\rm eff}(\phi)=V(\phi) + \rho_{\rm matter} e^{\phi/M}.
\end{equation}
As a result, even though $V$ might have a runaway form,  the conditions on $V(\phi)$ ensure that the effective
potential has a minimum at $\phi = \phi_{\rm min}(\rho_{\rm matter})$ where
\begin{equation}
V^{\prime}_{\rm eff}(\phi_{\rm min}) = 0 = V^{\prime}(\phi_{\rm min})
+ \frac{\rho_{\rm matter}}{M} e^{\phi_{\text{min}}/M}. 
\end{equation}

Whether or not the chameleon mechanism is both active and strong
enough to evade current experimental constraints depends partially on
the details of the theory, i.e. $V(\phi)$ and $M$, and partially on
the initial conditions (see Refs. \cite{chamKA,chamstrong,chamcos} for
a more detailed discussion). For exponential matter couplings and a
potential of the form
\begin{equation}\label{poti}
V(\phi)= \Lambda^4\exp (\Lambda^n/\phi^n) \approx \Lambda^4 +
\frac{\Lambda^{4+n}}{\phi^n},
\end{equation}
the chameleon mechanism can in principle hide the field such that
there is no conflict with current laboratory experiments, solar system
or cosmological observations \cite{chamKA,chamcos}. Importantly, for a
large range of values of $\Lambda$, the chameleon mechanism is strong
enough in such theories to allow even strongly coupled theories with
$M \ll M_{Pl}$ to have remained undetected \cite{chamstrong}. The
first term in $V(\phi)$ corresponds to an effective cosmological
constant whilst the second term is a Ratra-Peebles inverse power-law
potential \cite{ratra}. If one assumes that $\phi$ is additionally
responsible for late-time acceleration of the universe then one must
require $\Lambda\approx \Lambda_c \equiv (2.4\pm 0.1) \times
10^{-12}\GeV$.

Throughout the rest of this paper, it is our aim to remain as general
as possible and assume as little about the precise form of $V(\phi)$
as is necessary.  However, when we come to more detailed discussions
and make specific numerical predictions, it will be necessary to choose
a particular form for $V(\phi)$. In these situations, we assume that
$V(\phi)$ has the following form:
$$
V(\phi) = \Lambda_c^4\left(1+\frac{\Lambda^{n}}{\phi^n}\right).
$$ 
We do this not because this power-law form of $V$ is in any way
preferred or to be expected, but merely as it has been the most widely
studied in the literature and because is the simplest with which to
perform analytical calculations.  The power-law form is also useful as
an example as it displays, for different values of the $n$, many of
the features that we expect to see in more general chameleon
theories. We also note how the predictions of a theory with $V(\phi) =
\Lambda^4_c \exp(\Lambda^n/\phi^n)$ differ from those of a theory with
a true power-law potential. With this choice of potential, the
constant term in $V(\phi)$ is responsible for the late time
acceleration of the Universe.

\section{Chameleon trapping and photon afterglow}\label{sec:chamjar}
For a chameleon-like scalar field, the classical field equations
following from \Eqref{action} are
\begin{eqnarray}
\square \mathbf{a} &=& \frac{\nabla \phi \times \mathbf{B}}{M}, \\
\square \phi - m^2 \phi &=& \frac{\mathbf{B} \cdot (\nabla \times \mathbf{a})}{M},
\end{eqnarray}
where we have used the Lorenz-gauge condition.  We take $\mathbf{B} =
B \mathbf{e}_{x}$ as the background magnetic field, and $\mathbf{a} =
a_{\parallel}\mathbf{e}_{x} + a_{\perp}\mathbf{e}_{y}$ as the
propagating photon, moving in the positive $z$ direction (``to the
right''). We perform a Fourier transform with respect to time,
\begin{eqnarray}
a_{\perp}(t,z) &=& \int \,\dd \omega \,a(\omega,z) \,e^{-i\omega t}, \\
\phi(t,z) &=&  -i \int \,\dd \omega \,\chi(\omega,z) \,e^{-i\omega t},
\end{eqnarray}
where we have dropped the label $\perp$, since the photon component
$a_\parallel$ parallel to the magnetic field anyway does not interact
with the chameleon at all. The notation $\chi(\omega,z)=i
\phi(\omega,z)$ is introduced here for later convenience.  Defining
$\tilde{a}(\omega,k)$ and $\tilde{\chi}(\omega,k)$ as the Fourier
transforms w.r.t. $z$ of $a(\omega,z)$ and $\chi(\omega,z)$, we
arrive at
\begin{eqnarray}
(\omega^2 - k^2)\tilde{a} &=& -\frac{B k}{M} \tilde{\chi}, \\
(\omega^2 - k^2 - m^2)\tilde{\chi} &=& -\frac{Bk}{M} \tilde{a}.
\end{eqnarray}
Solutions exist if
$$
(\omega^2 - k^2 - m^2)(\omega^2 - k^2) = \frac{B^2 k^2}{M^2},
$$
the roots of which define the dispersion relations,
\begin{equation}
k_{\pm}^2 = \omega^2 - \left(m^2 -
\frac{B^2}{M^2}\right)\left(\frac{\cos 2 \theta \pm 1}{2\cos
  2\theta}\right), 
\end{equation}
where
\begin{equation}
\tan 2\theta = \frac{2\omega B}{M\left(m^2 - \frac{B^2}{M^2}\right)}.
\label{theta}
\end{equation}
Defining $k_{\pm}=+\sqrt{k^2_{\pm}}$, the general solutions $a$ and
$\chi$ for the equations of motion read:
\begin{eqnarray}
a(\omega, z) &=&  a_{r}^{-}(\omega) e^{ik_{-} z} + \tan^2 \theta
   a_{r}^{+}(\omega)e^{i k_{+} z} + a_{l}^{-}(\omega)e^{-ik_{-} z} +
   \tan^2 \theta a_{l}^{+}(\omega)e^{-i k_{+} z}, \label{eom1}\\ 
\chi(\omega, z) &=& \frac{\omega}{k_{-}} \tan \theta
   \left(a_{r}^{-}(\omega) e^{ik_{-} z}-a_{l}^{-}(\omega)e^{-ik_{-}
     z}\right)\nonumber\\ 
&&-\frac{\omega}{k_{+}} \tan \theta
   \left(a_{r}^{+}(\omega) e^{ik_{+} z}-a_{l}^{+}(\omega)e^{-ik_{+}
     z}\right), \label{eom2}
\end{eqnarray}
where $a_l (a_r)$ is the amplitude of the wave traveling to the left (right).
So far, the above equations are very similar to those of a laser
interaction with a scalar ALP (or dilaton-like particle) in a magnetic
field \cite{Maiani:1986md}. The important difference between a scalar ALP and a
chameleon is due to the boundary conditions at the ends of the optical
vacuum chamber: whereas an ALP is considered to be weakly interacting, the
chameleon is reflected at the chamber ends and thus ``trapped'' in the vacuum chamber. 

We begin by considering the simplest set-up for an analytic study,
wherein the two ends of the vacuum chamber (``jar'') are located right at edge
of the magnetic interaction region, i.e., $B > 0$ inside the jar and
$B=0$ outside. We also confine ourselves to an experiment where the
photon field is not stored in an optical cavity as for ALP searches,
but simply enters, passes through and leaves the interaction region.
The chameleon field is however trapped between
two optical windows of the vacuum chamber.

The chameleon field is taken to reflect perfectly off the walls of the
jar which are located at $z=L$ and $z=0$, whereas the photons only
enter the jar at $z=0$ and pass straight through. The reflection
of the chameleon field implies that $\partial_{z}\chi = 0$ at $z=0$
and $z=L$. This gives
\begin{eqnarray}
a_{r}^{-} + a_{l}^{-} &=& a_{r}^{+} + a_{l}^{+}, \label{c1}\\
a_{r}^{-}e^{ik_{-}L} + a_{l}^{-}e^{-ik_{-}L} &=&
a_{r}^{+}e^{ik_{+}L}+a_{l}^{+}e^{-ik_{+}L}.\label{c2} 
\end{eqnarray}
For the photon boundary conditions, it is useful to introduce the operators
$\mathcal{R}=\omega -i \partial_z$ and $\mathcal L = \omega+ i
\partial_z$ which project onto right- and left-moving photon
components in vacuum. The condition that no photons enter the jar on
the right side at $z=L$, $\mathcal L a(\omega,z)|_{z=L}=0$, gives: 
\begin{eqnarray}
&&a_{r}^{-}\left(1-\frac{k_{-}}{\omega}\right)e^{ik_{-}L} + \tan^2
  \theta a_{r}^{+}\left(1-\frac{k_{+}}{\omega}\right)e^{ik_{+}L}
+  a_{l}^{-}\left(1+\frac{k_{-}}{\omega}\right)e^{-ik_{-}L}+\tan^2
  \theta
  a_{l}^{+}\left(1+\frac{k_{+}}{\omega}\right)e^{-ik_{+}L}=0. \quad
  \label{c3}
\end{eqnarray}
Let as assume that the photon field entering the jar at $z=0$ has the
form
\begin{equation}
a_{\text{in}}(\omega,z\leq 0) = \alpha(\omega)e^{i k z},\label{incond}
\end{equation}
with the vacuum dispersion relation $k=\omega$. Then, the photon boundary
condition at $z=0$ is given by $\mathcal R
a_{\text{in}}(\omega,z)|_{z=0}= \mathcal R a(\omega,z)|_{z=0}$,
yielding
\begin{equation}
2\alpha = a_{r}^{-}\left(1+\frac{k_{-}}{\omega}\right) + \tan^2 \theta
a_{r}^{+}\left(1+\frac{k_{+}}{\omega}\right) +
a_{l}^{-}\left(1-\frac{k_{-}}{\omega}\right)+\tan^2 \theta
a_{l}^{+}\left(1-\frac{k_{+}}{\omega}\right).\label{in}
\end{equation}
Equations \eqref{c1}-\eqref{c3} and \eqref{in} determine the
photon amplitudes $a_{l,r}^{\pm}$ completely, and a full solution is
straightforward. The physical signature of the chameleon field is
encoded in the outgoing photon that leaves the jar at $z=L$ and which we
parametrize as
$$
a_{\text{out}}(\omega,z\geq L) = \beta(\omega)e^{i k z},
$$
again with the vacuum dispersion $k=\omega$. The form of the wave
packet $\beta(\omega)$ as a function of the amplitudes $a_{r,l}^\pm$
is determined by the matching condition $\mathcal R
a(\omega,z)|_{z=L}= \mathcal R a_{\text{out}}(\omega,z)|_{z=L}$,
implying
\begin{equation}
2\beta = a_{r}^{-}\left(1+\frac{k_{-}}{\omega}\right)e^{ik_{-}L} +
\tan^2 \theta a_{r}^{+}\left(1+\frac{k_{+}}{\omega}\right)e^{ik_{+}L}
+ a_{l}^{-}\left(1-\frac{k_{-}}{\omega}\right)e^{-ik_{-}L}+\tan^2
\theta
a_{l}^{+}\left(1-\frac{k_{+}}{\omega}\right)e^{-ik_{+}L}.\quad \label{bout}
\end{equation}
Since we expect $M$ to be a scale beyond the particle-physics standard
model, and $m= \mathcal{O}$(1~meV), the dimensionless
combination $B/(mM)$ can be considered as a very small parameter for all
presently conceivable laboratory field strengths. For typical
laboratory laser frequencies $\omega$, the whole right-hand side of
\Eqref{theta}  is small, implying that
\begin{equation}
\theta \simeq \frac{\omega B}{m^2M}
\end{equation}
is a small expansion parameter for the present problem. We also assume
that the laser frequency is larger than the vacuum chameleon mass,
$m^2/\omega^2\ll 1$, but the combination $m^2 L/\omega$ can still be
a sizable number owing to the length of the jar. In these limits,
where $k_{+} \approx \omega - m^2/2\omega$ and $k_{-} \approx \omega
$, the outgoing wave packet reduces to
\begin{equation}
\beta(\omega) \approx e^{i\omega L}\alpha(\omega) \left[ 1 +  2i
  \theta^2 \left( \frac{m^2 L}{4\omega} - 
  \frac{\sin\left(\frac{m^2 L}{4\omega}\right)\sin\left(\omega L -
    \frac{m^2 L}{4\omega}\right)}{\sin\left(\omega L - \frac{m^2
      L}{2\omega}\right)}\right) \right]. 
\end{equation}
Let us study the quotient
$$
Q \equiv - 2i\frac{\sin\left(\frac{m^2
    L}{4\omega}\right)\sin\left(\omega L - \frac{m^2
    L}{4\omega}\right)}{\sin\left(\omega L - \frac{m^2
    L}{2\omega}\right)}
= \left(e^{-i \frac{m^2L}{2\omega}}-1\right) -
   2\left(1-\cos({\scriptstyle \frac{m^2L}{2\omega}})\right)
   \sum_{n=1}^{\infty} e^{2 i n(\omega L
  - \frac{m^2L}{2\omega})}. 
$$
We assume that the photon wave packet that is sent into the jar
$\alpha(\omega)$ is strongly peaked about $\omega =
\bar{\omega}$. Close to $\bar{\omega}$, we then have
\begin{eqnarray*}
Q& \approx& \left(e^{-\frac{i m^2 L}{\bar{\omega}}}e^{i\omega
  \frac{m^2 L}{2\bar{\omega}^2}}-1\right) \nonumber\\
&& +
  \sum_{n=1}^{\infty}\left[e^{-\frac{(2n+1)i m^2
  L}{\bar{\omega}}}e^{i\omega(2n 
  L + (2n+1)\frac{m^2 L}{2\bar{\omega}^2})}  +
  e^{-\frac{(2n-1)i m^2 L}{\bar{\omega}}}e^{i\omega(2n L +
  (2n-1)\frac{m^2 L}{2\bar{\omega}^2})}
  -2e^{-\frac{2in m^2 L}{\bar{\omega}}}e^{i\omega(2n
  L + n m^2 L /\bar{\omega}^2)}\right]. 
\end{eqnarray*}
Fourier transforming back to real time, the outgoing photon field can
be expressed by the functional form of the ingoing photon as:
\begin{eqnarray}
a_{\text{out}}(t) &\approx& a_{\text{in}}\left(t-L-\frac{B^2 L}{2M^2
     m^2}\right) 
  + \frac{B^2}{M^2 m^4}a_{\text{in}}^{\prime \prime}(t-L) 
  - \frac{B^2}{M^2 m^4} e^{-i \frac{m^2 L}{\bar{\omega}}}
     a_{\text{in}}^{\prime \prime}\left(t-L-\frac{m^2
     L}{2\bar{\omega}^2}\right) 
   \nonumber\\ 
&&- \frac{B^2}{M^2  m^4}\sum_{n=1}^{\infty}
   \left[e^{-i\frac{(2n+1)m^2 L}{\bar{\omega}}}
  a_{\text{in}}^{\prime \prime}\left(t-(2n+1)L - (2n+1)\frac{m^2
    L}{2\bar{\omega}^2}\right) \right. \nonumber\\ 
&&\left.\qquad\qquad 
  +  e^{-i\frac{(2n-1)m^2 L}{\bar{\omega}}} a_{\text{in}}^{\prime
    \prime}\left(t-(2n+1)L - (2n-1)\frac{m^2
    L}{2\bar{\omega}^2}\right)\right.\nonumber\\
&&\left.\qquad\qquad
  -2e^{-i\frac{2nm^2L}{\bar{\omega}}} a_{\text{in}}^{\prime
     \prime}\left(t-(2n+1)L - \frac{n m^2
     L}{\bar{\omega}^2}\right)\right] + O(\theta^4), 
\label{Aout1}
\end{eqnarray}
where $a_{\text{in}}^{\prime \prime} = \dd^2 a_{\text{in}}/\dd t^2$,
and we have suppressed the $z$ dependence which comes in the form of a
plane wave $e^{i k z}$. The terms in brackets $\left[\cdot\right]$
represent the afterglow effect for the $\bot$ photon mode.  Let us
assume that $a_{\text{in}} \propto e^{-i \bar{\omega} t}$ for $0 < t <
T$ and vanishes otherwise, and define $N=T/L$.  It is clear that
unless $N m^2 L /\bar{\omega} = T m^2 /\bar{\omega} \ll 1$ the
different contributions to the afterglow effect will generally interfere
destructively.  If $m^2 T /\bar{\omega} \gg 1$ and $N \gg 1$, the
afterglow effect will scale as $1/N$.  If $m \sim O(1{\rm meV})$ then
$m^2 L /\bar{\omega}$ can be of order $ O(1)$ and so one must ensure
that $T/L \sim O(1)$ or smaller for the afterglow effect not to be
affected by interference.  With $L \sim O(1{\rm m})$, this requires
$T$ to be no greater than a few nanoseconds.  The GammeV experiment
\cite{GammeV} uses $5{\rm ns}$ wide pulses which avoids interference
effects. On the other hand, it may be possible to exploit the
interference effects for increasing the sensitivity or a determination
of the chameleon mass; see the appendix.

Let us generalize the above result to the case, where the jar is
longer than the interaction region, as is, for instance, the case for
the GammeV experiment \cite{GammeV}. We let $z=0$ label the beginning of
the interaction region. The chameleon reflects off the jar at $z=-d$
and $z=L$. If $0 < z < L$, the solution to the equation of motions
Eqs.~\eqref{eom1} and \eqref{eom2} still hold.

Outside the interaction region but inside the jar for $-d\leq z\leq
0$, we have: 
\begin{eqnarray}
a(\omega,-d\leq z\leq0) &=&  a_r^d(\omega) e^{i\omega z} +
a_l^d(\omega) e^{-i\omega z}, \label{photd}\\ 
\chi(\omega,-d\leq z\leq 0) = &=& \tan \theta\left(c_{r}(\omega) e^{ik_{m}z} -
   c_{l}(\omega) e^{-ik_{m}z}\right), \label{chamd}
\end{eqnarray}
where $k_{m} \equiv \sqrt{\omega^2 - m^2}$ and the amplitudes
$a^d_{l,r}$ and $c_{l,r}$ need to be determined by boundary and
matching conditions. In analogy to \Eqref{incond}, we specify the
boundary condition of the ingoing wave packet as
\begin{equation}
a_{\text{in}}(\omega,z\leq -d) = \alpha(\omega)e^{i k (z+d)},\label{incondd}
\end{equation}
where $k=\omega$ is the vacuum dispersion. Matching the photon
amplitudes at the left end of the jar, $z=-d$, where the ingoing wave
is purely right-moving, $\mathcal R a_{\text{in}}(\omega,z)|_{z=-d} =
\mathcal R a(\omega,z)|_{z=-d}$, fixes the amplitude $a_r^d(\omega)=
\alpha(\omega) e^{i kd}$ of \Eqref{photd}. Matching the photon
amplitudes of \Eqref{photd} and \Eqref{eom1} at $z=0$ gives  
\begin{equation}
2 e^{ikd} \alpha= a_{r}^{-}\left(1+\frac{k_{-}}{\omega}\right) + \tan^2 \theta
a_{r}^{+}\left(1+\frac{k_{+}}{\omega}\right) +
a_{l}^{-}\left(1-\frac{k_{-}}{\omega}\right)+\tan^2 \theta
a_{l}^{+}\left(1-\frac{k_{+}}{\omega}\right).\label{ind}
\end{equation}
for the right-movers. The corresponding left-mover equation fixes the
amplitude $a_l^d(\omega)$ of \Eqref{photd} which contains information
about the afterglow effect at the left end of the jar. Here, we
concentrate on the afterglow at the right end. For the matching of
the chameleon field at $z=0$, we act with the massive left- and
right-moving projectors $\mathcal{L}_m=(k_m+i \partial_z)$,
$\mathcal{R}_m=(k_m-i \partial_z)$ on Eqs.~\eqref{eom2} and
\eqref{chamd}, yielding 
\begin{eqnarray}
c_{r} =
\frac{1}{2}\left(\frac{\omega}{k_{m}}+\frac{\omega}{k_{-}}\right)a_{r}^{-}
+ \frac{1}{2}\left(\frac{\omega}{k_{m}}-\frac{\omega}{k_{-}}\right)a_{l}^{-}
- \frac{1}{2}\left(\frac{\omega}{k_{m}}+\frac{\omega}{k_{+}}\right)a_{r}^{+}
- \frac{1}{2}\left(\frac{\omega}{k_{m}}-\frac{\omega}{k_{+}}\right)a_{l}^{+},
\label{crd}\\ 
c_{l} =
\frac{1}{2}\left(\frac{\omega}{k_{m}}-\frac{\omega}{k_{-}}\right)a_{r}^{-}
+ \frac{1}{2}\left(\frac{\omega}{k_{m}}+\frac{\omega}{k_{-}}\right)a_{l}^{-}
- \frac{1}{2}\left(\frac{\omega}{k_{m}}-\frac{\omega}{k_{+}}\right)a_{r}^{+}
-
\frac{1}{2}\left(\frac{\omega}{k_{m}}+\frac{\omega}{k_{+}}\right)a_{l}^{+}. 
\label{cld}
\end{eqnarray}
The reflection of the chameleon field at $z = -d$ is equivalent to
\begin{equation}
c_{r} = -c_{l}e^{2ik_{m}d}.\label{ccond}
\end{equation}
This replaces Eq.~(\ref{c1}) whilst Eqs. (\ref{c2}) and (\ref{c3})
still hold. In conclusion, the matching conditions
Eqs.~\eqref{c2},\eqref{c3} and \Eqref{ccond} (together with
Eqs.~\eqref{crd} and \eqref{cld}) and the boundary condition
\Eqref{ind} completely determine the photon amplitudes
$a_{r,l}^\pm(\omega)$. In the limit of small $\theta$ and small
$m^2/\omega^2$, the outgoing wave packet $\beta(\omega)$, which is
still given by \Eqref{bout}, can be expressed as
\begin{equation}
\beta(\omega) \approx e^{i\omega (L+d)}\alpha(\omega) +  2i \theta^2
e^{i \omega (L+d)} \alpha(\omega) \left[ \frac{m^2 L}{4\omega} -
  \frac{\sin\left(\frac{m^2 L}{4\omega}\right)\sin\left(\omega (L+d) -
    \frac{m^2 (L+2d)}{4\omega}\right)}{\sin\left(\omega (L+d) -
    \frac{m^2 (L+d)}{2\omega}\right)}\right].
\end{equation}
Again, for $a_{\text{in}}(t)$ being dominated by the frequency
$\bar{\omega}$ at $z=-d$, the outgoing wave $a_{\text{out}}(t)$ reads
\begin{eqnarray}
a_\text{{out}}(t) 
&\approx& a_{\text{in}}\left(t-L-d-\frac{B^2 L}{2M^2 m^2}\right)+
  \frac{B^2}{M^2 m^4} a_{\text{in}}^{\prime \prime}(t-L-d)
  - \frac{B^2}{M^2 m^4} e^{-i \frac{m^2 L}{\bar{\omega}}}
   a_{\text{in}}^{\prime \prime}\left(t-L-d-\frac{m^2
     L}{2\bar{\omega}^2}\right) \nonumber\\
&& - \frac{B^2}{M^2 m^4}\sum_{n=1}^{\infty}\left[e^{-i\frac{m^2
      ((2n+1)L+2n d)}{\bar{\omega}}} a_{\text{in}}^{\prime
    \prime}\left(t-(2n+1)(L+d)  - \frac{m^2
    ((2n+1)L+2nd)}{2\bar{\omega}^2}\right) \right. \nonumber\\
 &&\qquad\qquad\qquad +\left.e^{-i\frac{m^2 ((2n-1)L+2n d)}{\bar{\omega}}}
  a_{\text{in}}^{\prime \prime}\left(t-(2n+1)(L+d)  - \frac{m^2
    ((2n-1)L+2nd)}{2\bar{\omega}^2}\right)\right.\nonumber\\
&&\qquad\qquad\qquad \left.-2e^{-i\frac{2n m^2(L+ d)}{\bar{\omega}}} 
  a_{\text{in}}^{\prime \prime}\left(t-(2n+1)(L+d)  - \frac{m^2
    n(L+d)}{\bar{\omega}^2}\right)\right] + O(\theta^4).
  \label{Aout2}  
\end{eqnarray}
It is important that we mention that by ignoring higher-order terms, we have dropped
the information about the decay in the afterglow effect at late
times. Since only a finite amount of energy initially converted into
chameleon particles for a finite laser pulse, the afterglow effect
must eventually decay, the time-scale of which can straightforwardly be
estimated: The probability that a chameleon particle converts to a
photon as it passes through the interaction region is
\begin{equation}
{\cal P}_{\varphi \rightarrow \gamma} = 4\theta^2\sin^2
\left(\frac{m^2 L}{4 \omega}\right). \label{eq:prob}
\end{equation}
After a time $t = (2N+1)(L+d)$, the chameleon field, which had been
created by the initial passage of the photons, has moved $N$ times
back and forth through the interaction region, i.e., from $z=L$ to
$z=-d$ and then back to $z=L$ again.  Since any chameleon particle
that is converted into photons escapes, the energy in the chameleon
field after time $t = (2N+1)(L+d)$ is reduced by a factor
\begin{equation}
F(N)=(1-{\cal P}_{\varphi \rightarrow \gamma})^{2N}. \label{eq:exfac}
\end{equation}
We therefore define the half-life $t_{1/2}$ of the afterglow effect
by $t_{1/2} = (2N_{1/2} +1)(L+d)$ where $F(N_{1/2})=1/2$.  Given that
$\theta^2 \ll 1$, this gives:
$$
t_{1/2} \approx  \frac{(L+d)\ln 2}{4\theta^2\sin^2 \left(\frac{m^2
    L}{4 \omega}\right)}. 
$$
We define $N_{\text{pass}}=2N_{1/2}$ to be the approximate number of
complete passes through the interaction region,
$$
N_{\text{pass}} = 2N_{1/2} \approx \frac{\ln 2}{4\theta^2\sin^2
  \left(\frac{m^2 L}{4 \omega}\right)}. 
$$
For a realistic estimate of the outgoing wave, we should then replace
the infinite upper limit of the sums in Eqs. (\ref{Aout1}) and
(\ref{Aout2}) by $N_{\text{pass}}$.  In the limit $m^2 L/4 \omega \ll
1$, we obtain
$$
t_{1/2} \approx \frac{4M^2(L+d)\ln 2}{B^2 L^2}.
$$
We observe that the dependence on the frequency $\omega$ and on the
chameleon mass $m$ drops out in this limit.  For a typical scale
$M\approx 10^{6}\,{\rm GeV}$ and experimental parameters $B \approx
5\,{\rm T}$ and $L+d \approx L = 6\,{\rm m}$, we obtain $t_{1/2} = 63\,{\rm
s}$ and $N_{\text{pass}} = 3 \times 10^{9}$, corresponding to a
time scale that should allow for a high detection efficiency.

\section{Signatures of the afterglow}\label{sec:sigaglow}

\subsection{General Signatures and Example I: the GammeV set-up}
Let us consider the simple case where the pulse length $T < 2(L+d)$,
such that chameleon and photon afterglow propagation inside the cavity
happens in well separated bunches. For a $5\,{\rm ns}$ pulse, this
corresponds to $(L+d) > 0.75{\rm m}$ which is, for instance, the case
in the GammeV experiment \cite{GammeV}.  As a consequence, there will be no interference
between the chameleonic afterglow and the initial pulse.  We assume
that for $0 < t < T$, $a_{\text{in}} = a_{0} e^{-i \bar{\omega} t}$,
and $a_{\text{in}}\simeq0$ before and after this time period.
Henceforth, we set $\bar{\omega} = \omega$. 

Then, the afterglow photons also come in bunches of time duration
$\approx T$. The $N$th bunch leaves the vacuum chamber to the right at
a time $t$ in the interval $(2N+1) (L+d) \lesssim t \lesssim
T+(2N+1)(L+d) < (2(N+1)+1)(L+d)$. For $1 \leq N \ll N_{1/2}$ and
defining $\tau = t - (2N+1)(L+d)$, the afterglow amplitude reads:
\begin{eqnarray}
a_{\text{out}}(t)e^{i\omega \tau} &=& -\frac{2B^2 \omega^2}{M^2 m^4}
\left(1-\cos\left(\frac{m^2 L}{2\omega}\right)\right) e^{-iN \frac{m^2
(d+L)}{\omega}}a_{0}, \\ &=& -\frac{4B^2 \omega^2}{M^2 m^4}
\sin^2\left(\frac{m^2 L}{4\omega}\right) e^{-iN \frac{m^2
(d+L)}{\omega}} a_{0}. \nonumber
\end{eqnarray}
We identify the modulus of the probability amplitude for an initial
photon to reappear in the $N$th afterglow pulse as
\begin{equation}
\mathcal P =  \frac{4B^2 \omega^2}{M^2 m^4} \sin^2\left(\frac{m^2
  L}{4\omega}\right).
\label{eq:P}
\end{equation}
Incidentally, this probability amplitude is identical to the
chameleon-photon conversion probability stated in \Eqref{eq:prob},
since the pulse-to-afterglow photon conversion involves a
photon-chameleon conversion twice.  Thus if each pulse contained
$n_{\gamma}^{\text{pulse}}$ photons, the number of afterglow photons produced
by this pulse after the time $(2N+1) (L+d) \lesssim t \lesssim
T+(2N+1)(L+d) < (2(N+1)+1)(L+d)$ with $1 \leq N \ll N_{1/2}$ is:
$$
n_{\gamma}^{\text{glow}}(N) \approx \mathcal{P}^2  n_{\gamma}^{\text{pulse}}.
$$
So far, we have neglected the afterglow decay at late times. For
larger values of $N$, we must account for the fact that the chameleon
amplitude decreases during the photonic afterglow. This is taken care
of by the extinction factor of \Eqref{eq:exfac},
\begin{eqnarray}
n_{\gamma}^{\text{glow}}(N) \approx \mathcal{P}^2
(1-\mathcal{P})^{2(N-1)} n_{\gamma}^{\text{pulse}}. 
\end{eqnarray}
For an initial laser pulse with energy $E_{\text{pulse}}$ and
frequency $\omega$, the number of photons is $n^{\text{pulse}}_\gamma=
E_{\text{pulse}}/\omega$. A characteristic quantity is given by the
detection rate of afterglow photons in the $N$th bunch,
\begin{equation}
R_{N} =  \mathcal{P}^2(1-\mathcal{P})^{2(N-1)}
\frac{E_{\text{pulse}}}{\omega T} \eta_{\text{det}}, 
\label{eq:RN}
\end{equation}
where $\eta_{\text{det}}\leq 1$ is the efficiency of the detector.  As
an example, let us consider the GammeV experiment \cite{GammeV} with a laser of
$T=5$~ns pulse duration, $E_{\text{pulse}}=160$~mJ pulse energy and
wave length $\lambda = 532$~nm.  This corresponds to an initial photon rate of
$E_{\text{pulse}}/\omega T = 8.8 \times 10^{25}\,{\rm s}^{-1}$. With a
magnetic field of $B = 5\,{\rm T}$ and length $L=6\,{\rm m}$ and
assuming that $m^2 L/4\omega \ll 1$, the early afterglow bunches
($N\ll N_{1/2}$) arrive at a rate of
$$
\frac{R_{N}}{\eta_{\text{det}}} \approx 2 \left(\frac{10^6\,{\rm
    GeV}}{M}\right)^4 \times 10^{-2} \,\,{\rm photons}\,\,{\rm
  pulse}^{-1}. 
$$
In the GammeV experiment \cite{GammeV}, the half-life, $t_{1/2}$, of the decay of the afterglow effect, in the limit $m^2 L/4\omega \ll 1$ and $L+d \approx L$, is:
\begin{equation}
\text{GammeV:}\quad  t_{1/2} \approx \frac{4M^2(L+d)\ln 2}{B^2 L^2} = 62.9\,{\rm s}\left(\frac{M}{10^{6}{\rm GeV}}\right)^2.
\end{equation}
The total number of photons contained in the afterglow after a single
pulse within the first half-life period $T < t < T+ t_{1/2}$ is
\begin{eqnarray}
n_{\gamma}^{\text{glow}}(t_{1/2}) &=& \sum_{j=1}^{N_{1/2}}
n_{\gamma}^{\text{glow}}(j) = \sum_{i=0}^{N_{1/2}-1}\mathcal{P}^2
(1-\mathcal{P})^{2i} n_{\gamma}^{\text{pulse}}  \nonumber\\
&=& \frac{\mathcal{P}^2 (1-(1-\mathcal{P})^{2N_{1/2}})}{(1-(1-\mathcal{P})^{2})} n_{\gamma}^{\text{pulse}} \approx \frac{\mathcal{P}}{4} n_{\gamma}^{\text{pulse}}\label{eq:41} , 
\end{eqnarray}
where we have used the definition of $N_{1/2}$ i.e. $(1-\mathcal{P})^{2N_{1/2}}=\frac{1}{2}$.
  In the limit of $m^2
L/4\omega \ll 1$, the number of afterglow photons in the first
half-life period, for instance, for the GammeV \cite{GammeV} experiment yields
$$
\text{GammeV:}\quad 
n_{\gamma}^{\text{glow}}(t_{1/2}) = \frac{B^2
  L^2}{16 M^2}n_{\gamma}^{\text{pulse}} \approx 2.5 \times  10^{7}
\left(\frac{10^{6}\,{\rm GeV}}{M}\right)^2. 
$$
With the conservative assumption of a detector efficiency of
$\eta_{\text{det}}\simeq 0.1$, the non-observation of any photon
afterglow from one laser pulse in the time $t < t_{1/2}$ would correspond to a lower bound on the
coupling scale $M> 1.6 \times 10^9$GeV in the regime of chameleon
(vacuum) masses $m<0.5$meV. However, to actually achieve such a bound one would have to run the experiment for at least a time of $t_{1/2}$ which for $M \sim 10^{9}$GeV is about $2$yrs.  It is therefore more practical to consider what lower bound on $M$ would result from the non-detection of any photons due to the afterglow of a single laser pulse after a time $t_{\rm expt} \ll t_{1/2}$.   We find:
$$
\text{GammeV:}\quad 
n_{\gamma}^{\text{glow}}(t_{\rm expt}) = \frac{B^4
  L^4}{16 M^4}n_{\gamma}^{\text{pulse}} \left(\frac{t_{\rm expt}}{2L}\right) \approx 3.1 \times 10^{4}
\left(\frac{10^{7}\,{\rm GeV}}{M}\right)^4 \left(\frac{t_{\rm expt}}{1\,{\rm min}}\right); \quad t_{\rm expt} \ll t_{1/2}. 
$$
With a minutes worth of measurement, the non-detection of any afterglow from a single pulse of the laser would for $m < 0.5$meV correspond to a lower bound on the coupling of $M > 7.5 \times 10^{7}\,{\rm GeV}$. If data were collected for a day then this constraint could be raised to $M > 4.6 \times 10^{8}\,{\rm GeV}$.

This should be read side by side with the currently best laboratory
bound for similar parameters for weakly coupled scalars or
pseudo-scalars derived from the PVLAS experiment
\cite{Zavattini:2007ee}; from the PVLAS exclusion limits for
magnetically-induced rotation at $B=2.3$Tesla, we infer $M\gtrsim
2\times 10^{6}$GeV.\footnote{Similar bounds have also been found by the BMV
experiment \cite{Robilliard:2007bq} and by the earlier BFRT experiment
\cite{Cameron:1993mr} from photon regeneration searches; however,
these bounds to not apply to chameleon models, since they involve a
passage of the hypothetical particles through a solid wall.} We
concluded that afterglow experiments are well suited for exploring
unknown regions in the parameter space of chameleon models.

As a side remark and a check of the calculation, we note that the total
number of photons contained in the full afterglow can be obtained from
\Eqref{eq:41} by extending the upper bound of the sum to infinity;
this yields
$$
n_{\gamma}^{\text{glow}}(t = \infty) =  \frac{1}{2}
\mathcal{P}n_{\gamma}^{\text{pulse}}, 
$$ 
which is one half of the total number of photons that would have
been initially converted into chameleon particles; the other half
creates an afterglow on the other (left) side of the vacuum chamber.

\subsection{Example II: An Optimized Experimental set-up}
As a further example, let us consider a more optimized experimental
set-up that nevertheless involves only parameters which are achievable
by current standard means or systems available in the near future. Our
optimized experimental set-up consists of an $L=14.3$m long magnetic
field of strength $B=9.5$T which corresponds exactly to that of the
OSQAR experiment at CERN \cite{OSQAR}. We consider a laser system
delivering a sufficiently short pulse with energy
$E_{\text{pulse}}\simeq 1$kJ and wavelength $\lambda=1053$nm. These
parameters agree with the specifications of the laser used by the BMV
collaboration \cite{Robilliard:2007bq} or of the PHELIX laser
\cite{PHELIX} which is currently being built at the GSI (with possible
upgrades to several kJ); its designed pulse duration is $1-10$ns,
corresponding to a required chamber length of $L+d>0.15-1.5$m. In
order to reduce the laser-beam energy deposit on the components of the
set-up, the beam diameter may simply be kept comparatively wide
(matching the geometry of the detector). Also the pulse duration could
be increased while keeping $E_{\text{pulse}}$ fixed. By increasing the
length $d$ of the non-magnetized part of the vacuum chamber, the pulse
duration can also be extended; e.g., choosing $d\simeq 100$m, the
pulse duration could be extended up to 1$\mu$s without disturbing
interference effects. Assuming $m^2 L/4\omega \lesssim \pi/2$, The
number of afterglow photons in the first half-life period for this
optimized set-up (i.e. $T=1\mu$s, $d=100$m, $B=9.5$T and $L=14.3$m)
yields
\begin{equation}
\text{optimized set-up:}\quad 
n_{\gamma}^{\text{glow}}(t_{1/2})  \approx 1.1 \times  10^{4}
\left(\frac{10^{10}\,{\rm GeV}}{M}\right)^2. 
\end{equation}
The half-life of the afterglow effect in this set-up is:
\begin{equation}
\text{optimized set-up:}\quad t_{1/2} \approx 58.5\,{\rm s} \left(\frac{M}{10^{6}\,{\rm GeV}}\right)^2
\end{equation}
Assuming that single photons can be detected in the optimized set-up,
$\eta_{\text{det}}=1$, the non-observation of any photon afterglow
during from one pulse in $t< t_{1/2}$ would correspond to a lower bound on the coupling
scale $M> 10^{12}$GeV in the regime of chameleon vacuum masses
$m<0.2$meV; however for such large values of $M$, the half-life of the afterglow is exceedingly large $t_{1/2} \sim 10^{6}\,{\rm yrs}$.  It is more reasonable then to ask what lower-bound one could place on $M$ if no photons are detected from the afterglow of a single pulse after a time $t_{\rm expt} \ll t_{1/2}$ has passed.  We find:
\begin{equation}
\text{optimized set-up:}\quad 
n_{\gamma}^{\text{glow}}(t_{\rm expt})  \approx 8.4
\left(\frac{10^{9}\,{\rm GeV}}{M}\right)^4 \left(\frac{t_{\rm expt}}{1\,{\rm min}}\right); \quad t_{\rm expt} \ll t_{1/2}. 
\end{equation}
We see that after one minute of measurements one could bound $M > 1.7
\times 10^{9}\,{\rm GeV}$, and if measurements could be conducted
constantly over a 24 hour period then this could be extended to $M >
1.0 \times 10^{10}\,{\rm GeV}$.

It is interesting to confront these potential afterglow laboratory
bounds with typical sensitivity scales obtained from astrophysical
arguments: the non-observation of solar axion-like particles by the
CAST experiment imposes a bound of $M>1.1 \times 10^{10}$GeV for
$m\lesssim 0.02$eV \cite{Andriamonje:2007ew}; slightly less stringent
bounds follow from energy-loss arguments for HB stars
\cite{Raffelt:2006cw}.  The similar order of magnitude demonstrates
that laboratory measurements of afterglow phenomena can probe scales
of new physics that have so far been explored only with astrophysical
observations. Of course, we need to stress that these numbers should
not literally be compared to each other, since they apply to different
theoretical settings; in particular, the astrophysical constraints do
not apply to chameleonic ALPs \cite{chamPVLAS,chamPVLASlong}, whereas
the non-observation of an afterglow would not constrain axion models.

\subsection{Example III: The BMV set-up}
As a final example, we consider the recent
light-shining-through-a-wall experiment performed by the BMV
collaboration \cite{Robilliard:2007bq}, where a laser with frequency
$\omega = 1.17{\rm eV}$, $T=4.8$~ns and $E_{\text{pulse}} \gtrsim
1$~kJ were used.  The magnetic field strength was $B=12.2$~T, but the
magnetic field remained at its maximum values for about
$150{\mu}\text{s}$; $L=2\times 0.45\,{\rm m}$.  If this set-up were
modified to search for chameleons trapped in a vacuum chamber then
(when $m^2 L/4\omega \ll 1$) we would find:
$$
\frac{R_{N}}{\eta_{\text{det}}} \approx 4.53 \left(\frac{10^6\,{\rm
    GeV}}{M}\right)^4 \,\,{\rm photons}\,\,{\rm pulse}^{-1}. 
$$
After $150\,\mu{\rm s}$ the chameleons will have made $\approx 25000$
complete passes through the vacuum chamber.  If $\theta \lesssim
10^{-3}$ then $t_{1/2} > 150\,\mu{\rm s}$, and so the duration of the
afterglow effect would be limited by the length of the magnetic pulse.
For $\theta \lesssim 10^{-3}$, the total number of afterglow photons
produced by each pulse that could potentially be detected in the time
interval $T < t < 150\,\mu{\rm s}$ is (given $\eta_{\rm det} \approx
0.5$ \cite{Robilliard:2007bq}):
$$
n_{\gamma}^{\text{glow}}(150\,\mu{\rm s}) \approx 5.7 \times
\left(\frac{10^7\,{\rm GeV}}{M}\right)^2\,\,{\rm pulse}^{-1}. 
$$
We conclude that also the BMV apparatus could search for a coupling
parameter beyond $M\simeq 10^7$GeV. Of course, all experimental
set-ups considered here can push the detection limits even further by accumulating
statistics. 

\subsection{Experimental Bounds on Chameleon Models}

\begin{figure}[tbp]
\centering
\includegraphics[scale=0.6] {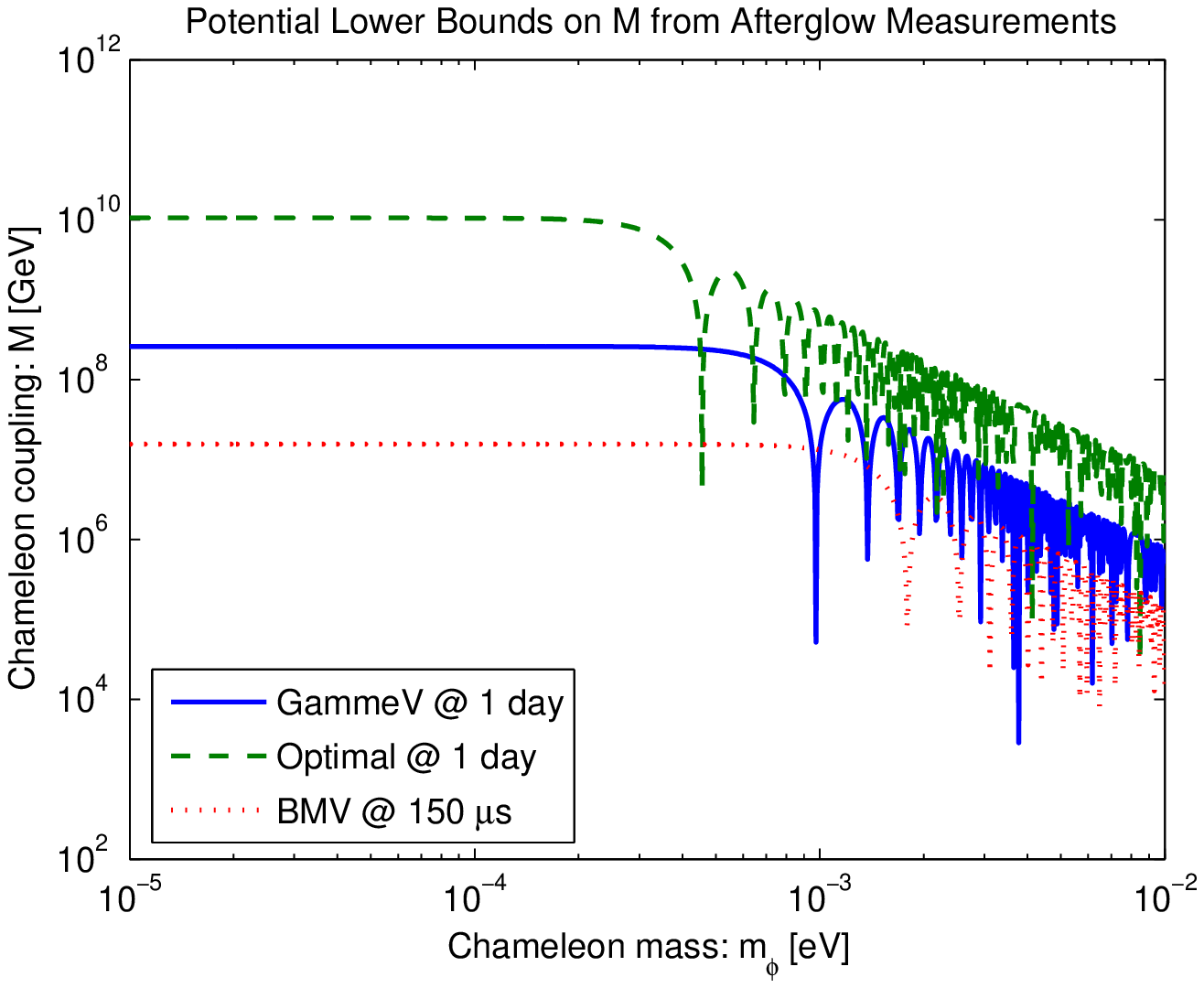}
\includegraphics[scale=0.6] {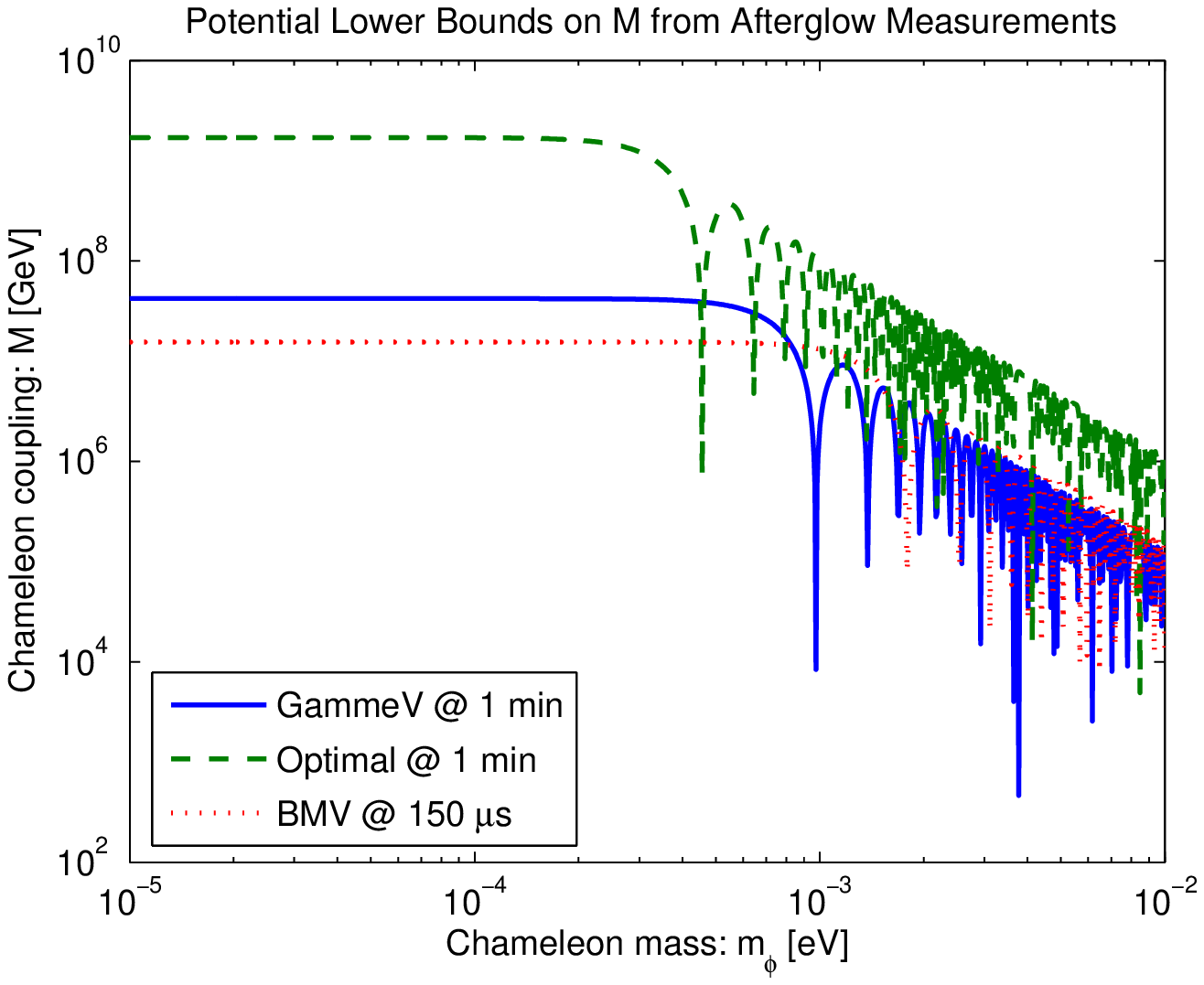}
\includegraphics[scale=0.6] {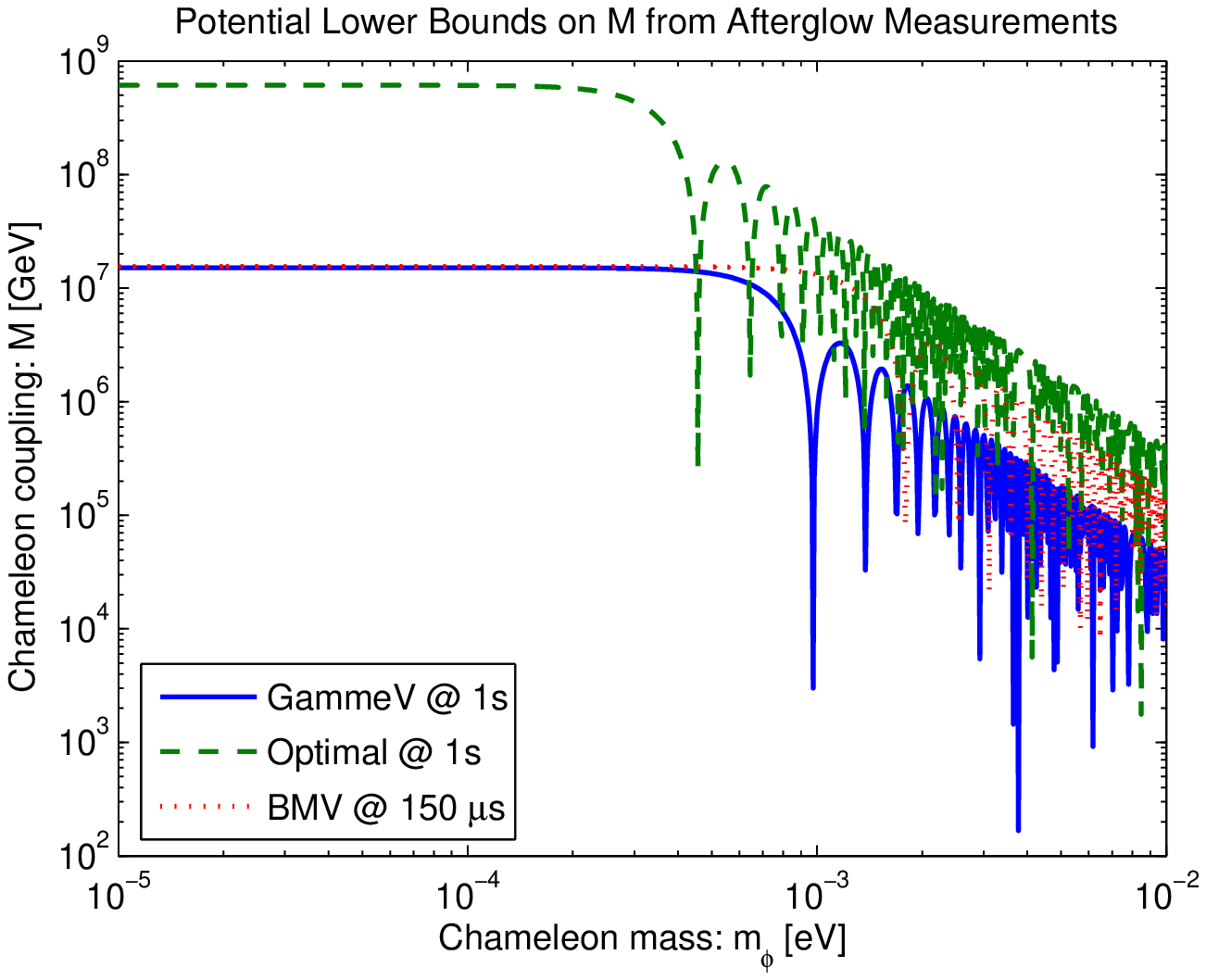}
\caption{Sensitivity limits for the scale $M$ specifying the inverse
chameleon-photon coupling vs. the chameleon mass for the BMV
\cite{Robilliard:2007bq} (red dotted line) and GammeV \cite{GammeV}
(blue solid line) experiments as well as a hypothetical optimized
experimental set-up (green dashed line); The experimental parameter
values are detailed in the main text. The top left panel corresponds
to an afterglow measurement with a duration $t_{\text{exp}}$ of one
day, the top right panel corresponds to one minute and the lower
figure to one second. The gain in sensitivity is less than an order of
magnitude by waiting a day rather than a minute. }
\label{fig}
\end{figure}

The potential experimental bounds on chameleon models derived from the
three examples discussed above are shown in Fig.~\ref{fig}. For the
full analysis, we retained the dependence on $m$ by taking the full
dependence of $\mathcal P$ on $m^2L/(4\omega)$ into account,
cf. \Eqref{eq:P}. In the case of weak coupling, i.e., large values of
$M$, typical values for the half-life of the afterglow are much longer
than an experimentally feasible time of measurement. The achievable
sensitivity therefore depends on the time duration $t_{\text{expt}}$
of the afterglow measurement; for instance, the maximal sensitivity
for $M$ scales like $M_{\text{max}}\sim t^{1/4}$. In Fig.~\ref{fig},
we show sensitivity bounds for three different measurement durations,
$t_{\text{expt}}=1$s, 1min, 1day. 

For small $m$, we rediscover the $m$-independent sensitivity limits
which have been discussed analytically in the preceding
subsections. For larger $m$, we observe the typical oscillation
pattern with sensitivity holes which correspond to full 2$\pi$
oscillation phases of the photon-chameleon system. 

\section{Background and systematic effects}\label{sec:back}
\subsection{Standard Reflection of Photons}

So far, we have assumed that the optical windows forming the caps
of the jar are completely transparent.  In a real experiment, this
transparency will be less than perfect and there will be some reflection of photons.
For simplicity, we assume that all photons incident on the optical
windows are either reflected or transmitted, i.e., we ignore any
absorption.  We also assume, for simplicity, that the photons hit the optical
windows perpendicularly.  The coefficients of reflection and
transmission are then given by
\begin{equation}
T = \frac{16 n_{\rm cap}^2 n_{\rm vac}^2}{(n_{\rm cap}+n_{\rm vac})^4}, \qquad R = 1-T,
\end{equation}
where $n_{\rm cap}$ and $n_{\rm vac}$ are the indices of refraction
for the optical window and the laboratory vacuum, respectively;
typical values are $n_{\rm vac}=1$ and $n_{\rm cap} = 1.5$ which gives
$R \approx 0.078$.  For the time interval $(2N+1) (L+d) \lesssim t
\lesssim T+(2N+1)(L+d) < (2(N+1)+1)(L+d)$, the rate of afterglow
photons leaving the jar at $z=L$ is given by \Eqref{eq:RN}.  In this
time interval, the  photons which have been trapped owing to
standard reflection leave the jar at $z=L$ with a rate
$$
R_{\rm reflect}(N) = T R^{2(N-1)} \frac{E_{\rm pulse}}{\omega T} \, \eta_{\text{det}},
$$
where we have neglected a possible interplay between reflection and
chameleon trapping; this is justified as long as the time scales for
reflection trapping and chameleon trapping are very different.
We define $N_{r}$ by the requirement that $R_{\rm reflect}(N) \leq
R_{\rm glow}(N)$ for all $N \geq N_{r}$. We then find that
$$ 
N_{r} \approx 1+ \left[\frac{\ln \mathcal{P} + R/2}{\ln R +
\mathcal{P}}\right] \approx 1+\frac{\ln \mathcal{P}}{\ln R},
$$
where we have used $\mathcal{P}$,~$R \ll 1$.  Even if $\mathcal{P}
\approx 10^{-30}$ we still have $N_{r} \approx 27$; if $\mathcal{P}
\approx 10^{-12}$ then $N_r \approx 11$. The effect of the standard
reflected photons is negligible compared with the afterglow effect for
$N > N_{r}$; moreover, it is clear that generally $N_{r} \ll
N_{\text{pass}}$. For example: with the GammeV parameters ($T=5$~ns and
$L+d \approx 6$~m), the afterglow dominates after $440$~ns if
$\mathcal{P} \approx 10^{-12}$ or after about $1100$~ns if
$\mathcal{P} \approx 10^{-30}$; both of these time scales are
generally far smaller than both the half-life of the afterglow effect
and the run-time of the experiments. We conclude that reflection of
photons off the caps of the vacuum chamber will not effect the potential of such
experiments to detect any chameleonic afterglow.

\subsection{Extinction of Chameleon Particles by a Medium}

Before we can be sure that the half-life of the chameleon afterglow is, as calculated above, accurate we must consider whether scattering and absorption of chameleon particles by the matter in the laboratory
vacuum inside the jar places an important upper limit on the number of
passes that the chameleon field makes through the jar.  Roughly
speaking, if light couples to matter consisting of particles with mass
$m_{\text{p}}$ with a strength $e^2$, then the chameleon field couples
to it with a strength $m_{\text p}^2/ M^2$.  As a beam of light with
frequency $\omega$ travels a distance $L+d$ through a medium of free
particles with charge $\pm e$, mass $m_{\text p} \gg \omega$ and
number density $n_{\text p}$, Thompson scattering reduces its
intensity by a factor $\exp(-\Gamma_{\gamma} (L+d))$ where:
$$
\Gamma_{\gamma} = \frac{8\pi n_{\text p} e^4}{3 m^2_{\text p}}.
$$
Analogously, the intensity of the beam of chameleon particles traveling
within a medium of free particles with mass $m_{\text p} \gg \omega$ and number
density $n_{\text p}$ is reduced by a factor $\exp(-\Gamma_{\phi} (L+d))$,
where
$$
\Gamma_{\phi} = \frac{8\pi n_{\text p} m^2_{\text p}}{3 M^4}.
$$
Let us consider $\Gamma_{\phi}$ for a chameleon field propagating through a
laboratory vacuum with pressure $P_{\rm vac}$. For simplicity, we
assume that the vacuum contains only N${}_{2}$ molecules, which gives
$$
\Gamma_{\phi} = 1.97 \times 10^{-30} \left(\frac{10^6\,{\rm GeV}}{M}\right)^4
\left(\frac{P_{\rm vac}}{{\rm torr}}\right)\,{\rm m}^{-1}.
$$
After a time $(2N+1) (L+d) \lesssim t \lesssim T+(2N+1)(L+d) <
(2(N+1)+1)(L+d)$, the chameleon particles that were created by the
initial laser pulse have traveled on a distance $2N(L+d)$ and so
scattering has reduced the intensity of the chameleon particles by
$F_{\rm scat}(N)=\exp(-2N \Gamma_{\phi} (L+d))$.  We define $N_{\rm scat}$ by
$F_{\rm scat}(N_{\rm scat})=1/2$, i.e.,
$$
N_{\rm scat} = \frac{\ln 2}{2 \Gamma_{\phi} (L+d)} \approx 1.8 \times 10^{29}
\left(\frac{M}{10^6\,{\rm GeV}}\right)^4 \left(\frac{{\rm
torr}}{P_{\rm vac}}\right)\left(\frac{{\rm m}}{L+d}\right).
$$
The corresponding half-life due to scattering $t_{1/2}^{\rm scat}$
is thus given by
$$
t_{1/2}^{\rm scat} = (2N_{\rm scat}+1) (L+d) \approx 5.9 \times
10^{20}\,{\rm s}\left(\frac{M}{10^6\,{\rm GeV}}\right)^4
\left(\frac{{\rm torr}}{P_{\rm vac}}\right).
$$
For $m(L+d)/4 \omega \lesssim O(1)$, $P \lesssim O(1)\, {\rm torr}$,
$L+d \sim O(1)\,{\rm m}$ and $M \gtrsim 10^{6}\,{\rm GeV}$, is it
clear that
$$
t_{1/2}^{\rm scat} \gg t_{1/2}.
$$
We conclude that the scattering of chameleon particles by atoms in the
laboratory vacuum is negligible over the time scales of interest, $t
\lesssim t_{1/2}$. 

\subsection{Quality of the Vacuum}
We have found that the mass of the chameleon particles inside the interaction region 
plays a role 
in determining the rate of afterglow photon emission.  In $(2N+1) (L+d)
\lesssim t \lesssim T+(2N+1)(L+d) < 2((N+1)+1)(L+d)$ this rate was found to be
$$
\mathcal{P}^2(1-\mathcal{P})^{2(N-1)} \frac{E_{\rm pulse}}{\omega T},
$$
and that the half-life for this effect is:
\begin{equation}
t_{1/2} = \frac{(L+d) \ln 2}{\mathcal{P}},\quad
\text{where}\quad
\mathcal{P} = \frac{4B^2 \omega^2}{M^2 m^4} \sin^2 \frac{m^2 L}{4\omega}.\label{Prep}
\end{equation}
One of the key properties of chameleon fields is that the mass, $m$,
is not fixed but depends on the density environment, which is in turn
determined by the quality or pressure, $P_{\rm vac}$, of the vacuum. From Eq.(\ref{Prep}), it is clear that the probability $\mathcal{P}$, and hence the
afterglow photon rate is greatly suppressed, if $m^2 L /4\omega \gg
\pi/2$.  Additionally, if $m^2 L/4\omega \ll \pi/2$ then $\mathcal{P}$
is almost independent of $m$. The effective local matter density to
which the chameleon couples is $\rho_{\rm eff} = \rho_{\rm vac} +
B^2/2$ and only $\rho_{\rm vac}$ depends on $P_{\rm vac}$.  Thus
decreasing $\rho_{\rm vac}$ can only reduce $\rho_{\rm eff}$ and hence
$m$ so far. In all experiments, the smaller vacuum pressure always
increases the range of potentials and parameter space than can be
detected or ruled out; the downside is that making $P_{\rm vac}$
smaller inevitably increases costs.

Once $V(\phi)$ and $M$ are specified, it is always possible to find some value of $\rho_{\rm eff}$ such that $m^2 L/4 \omega = \pi/2$; we denote the special value of the effective ambient density by $\bar{\rho}_{\rm eff}$.  If $\bar{\rho}_{\rm eff} - B^2/2 > 0$, then $\rho = \bar{\rho}_{\rm eff}$ can be realized by setting $\rho_{\rm vac} = \bar{\rho}_{\rm vac} = \bar{\rho}_{\rm eff} - B^2/2$.  Since $m$ depends on $\rho_{\rm vac}$ only through $\rho_{\rm eff}=\rho_{\rm vac}+B^2/2$, there is little point in making $\rho_{\rm vac} \ll B^2/2$ as opposed to say $\rho_{\rm vac} \approx 0.1 B^2/2$, as it will increase costs but result in no great chance in the value of $\rho_{\rm eff}$.  We therefore define the critical value of the vacuum density as so:
$$
\rho^{\rm crit}_{\rm vac}(M,B) =\max\left(B^2/2, \bar{\rho}_{\rm eff} - B^2/2\right).
$$
The value of $\rho^{\rm crit}_{\rm vac}$ is significant as the
afterglow rate for $\rho_{\rm vac} \gg \rho_{\rm vac}^{\rm crit}(M,B)$
is greatly suppressed compared to the rate for $\rho_{\rm vac}
\lesssim \rho_{\rm vac}^{\rm crit}(M,B)$. There is also relatively
little gain in sensitivity in having $\rho_{\rm vac} \ll \rho_{\rm
vac}^{\rm crit}(M,B)$ as opposed to be $\rho_{\rm vac} \sim \rho_{\rm
vac}^{\rm crit}(M,B)$.  If one is interested in searching for
chameleon fields with $M \gtrsim M_{\rm min}$, where $M_{\rm min}$ is
the smallest value of $M$ in which one is interested (e.g. the
smallest value of $M$ that is not already ruled out by other
experiments), the optimal choice for the density of the vacuum, both
in terms of experimental sensitively and cost, is $\rho_{\rm vac} \sim
\rho_{\rm vac}^{\rm crit}(M_{\rm min},B)$.
   
For definiteness we consider a chameleon theory with power-law potential:
$$
V(\phi) = \Lambda_c^4(1+\Lambda^n/\phi^n)
$$
for some $n$ and with $\Lambda = \Lambda_c = 2.4\times 10^{-3}\,{\rm
eV}$.  The scale of $\Lambda$ could be seen as `natural' if the
chameleon field is additionally responsible for the late time
acceleration of the Universe. We use parameters for the GammeV
experiment \cite{GammeV} to provide an example.  In this set-up
$\omega = 2.3\,{\rm eV}$, $\Lambda =2.3 \times 10^{-3}\,{\rm eV}$,
$B=5$~T and $L = 6\,{\rm m}$. The most recent PVLAS results imply that
$M \gtrsim 10^{6}\,{\rm GeV}$ \cite{chamPVLAS}. Taking $M_{\rm min} =
10^{6}\,{\rm GeV}$, we find that with $B=5\,{\rm T}$:
\begin{eqnarray*}
\rho^{\rm crit}_{\rm vac}(n=1/2, M_{\rm min},B) &=& 3.2 \times 10^{-10}\,{\rm kg\,m}^{-3},\\
\rho^{\rm crit}_{\rm vac}(n=1, M_{\rm min},B) &=& 2.8 \times 10^{-10}\,{\rm kg\,m}^{-3}, \\
\rho^{\rm crit}_{\rm vac}(n=4, M_{\rm min},B) &=& 2.6 \times 10^{-11}\,{\rm kg\,m}^{-3}.
\end{eqnarray*}
If $\rho_{\rm vac} \lesssim 10^{-10}\,{\rm kg\,m}^{-3}$ then in this
set-up, $m^2 L/4\omega < \pi/2$ for $0.053 < n < 2.6$, and if
$\rho_{\rm vac} \lesssim 10^{-11}\,{\rm kg\,m}^{-3}$ then we have $m^2
L/4\omega < \pi/2$ for $0.017 < n < 4.4$.  For models with a power-law
potential, the best trade-off between cost and sensitivity for GammeV
set-up, would therefore be to use $\rho_{\rm vac} \approx 10^{-11} -
10^{-10}\,{\rm kg\,m}^{-3}$.  This corresponds to an optimal vacuum
pressure: $P_{\rm vac} \lesssim 10^{-8} - 10^{-7}\,{\rm torr}$.  A
vacuum of this quality was used in the recent PVLAS axion search
\cite{Zavattini:2005tm,Zavattini:2007ee}.  If one could rule out theories with $M <
10^{8}\,{\rm GeV}$ by some other means then very little would be
gained in terms of sensitivity to models with $n \sim O(1)$ by making
the vacuum pressure much smaller than: $P_{\rm vac}^{\rm crit} \sim
10^{-5}-10^{-4}\,{\rm torr}$.

If we consider instead the ``optimal set-up" that we described in
Section \ref{sec:sigaglow}, then for $M_{\rm min} = 10^{8}\,{\rm GeV}$
and $n \sim O(1)$, the optimal choice for the vacuum pressure is
$P_{\rm vac}^{\rm crit} \sim 10^{-6}-10^{-5}\,{\rm torr}$.  In the
context of chameleon models with a power-law potential with $n \sim
O(1)$ and $\Lambda \approx \Lambda_c = 2.4 \times 10^{-3}\,{\rm eV}$,
relatively little would be gained in terms of sensitivity by lowering
$P_{\rm vac}$ further than this.  However, lower values of $P_{\rm
vac}$ would increase the ranges of values of $\Lambda$ and $n$ that
could be detected.

Notice that, since one can vary the chameleon mass, $m_{\phi}$, by
changing the local density, $\rho_{\rm vac}$, it would in principle be possible
to measure $m_{\phi}$ using these experiments as a function of
$\rho_{\rm eff}$ by making use of the relation $\rho_{eff} = -MV^{\prime}(\phi)$. One would
then know $V^{\prime \prime}(\phi) \equiv m_{\phi}^2$ as a function of $V^{\prime}(\phi)$ and hence also $\phi(V^{\prime}(\phi))$ up to a constant. From this one could reconstruct $V^{\prime}(\phi+{\rm const})$. By integrating, one would then arrive at $V(\phi+const)+{\rm const}$, for some range of $\phi$. However, since the minimum size of $\rho_{\rm eff}$ is limited by $B^2/2
\approx 10^{-10} \text{kg\, m}^{-3}$ for $B = 5T$, one cannot probe $V(\phi)$
in the region that is cosmologically interesting today, i.e. $\rho \approx
10^{-27} \text{kg \, m}^{-3}$.  It must be stressed though that the
sensitivity of ALP experiments to $m$ is strongly peaked about values
for which $m^2 L/4 \omega \approx \pi/2$, and so it is not the most ideal
probe of $V(\phi)$; ALP experiments are best suited to measuring or placing lower-bounds on
$M$.  Casimir force measurements has recently be shown in Ref. \cite{chamcas} to provide a potentially much more
direct probe of $V(\phi)$ but conversely say little about the
chameleon to matter coupling, $M$.

\section{Conclusions}\label{sec:con}

In this paper, we have investigated the possibility of using an
\emph{afterglow} phenomenon as a unique chameleon trace in optical
experiments. The existence of this afterglow is directly linked with
the environment dependence of the chameleon mass parameter. The latter
causes the trapping of chameleons inside the vacuum chamber where they
have been produced, e.g., by a laser pulse interacting with a strong
magnetic field. The afterglow itself is linearly polarized
perpendicularly to the magnetic field.

We find that the trapping can be so efficient that the re-conversion of
chameleons into photons in a magnetic field causes an afterglow over
macroscopic time scales. For instance, for values of the inverse
chameleon-photon coupling $M$ slightly above the current detection
limit $M\sim 10^{6}$GeV \cite{Zavattini:2007ee} and magnetic fields
of a few Tesla and few meters long, the half-life of the afterglow is
on the order of minutes. Current experiments such as  ALPS, BMV,
GammeV and OSQAR can improve the current limit on $M$ by a few orders of
magnitude. With present-day technology, even a parameter range of
chameleon-photon couplings appears accessible which is comparable in
magnitude, e.g., to bounds on the axion-photon coupling derived from
astrophysical considerations, i.e., $M\sim 10^{10}$GeV. 

In the present work, we mainly considered the afterglow from an
initial short laser pulse, the associated length of which fits into
the optical path length within the vacuum chamber. From a technical
viewpoint, this choice avoids unwanted interference effects, but it
also has an experimental advantage: the resulting afterglow photons
all arrive in pulses of the same duration as the initial pulse and are
separated by a known time (i.e., $2(L+d)$). This can be useful in
extracting the signal from any background noise which should not be
correlated in this way. Also the polarization dependence of the
afterglow can be used to distinguish a possible signal from noise. 

As discussed in the appendix, also long laser pulses or continuous
laser light could be used in an experiment, if the resulting
interference can be controlled to some extent such that a
chameleon-photon resonance exists at least in some part of the
apparatus. This resonance phenomenon is particularly sensitive to
smaller chameleon masses. Unfortunately, a full control of the
chameleon resonance may experimentally be very difficult; but if it
were possible, the gain in afterglow photons, and subsequently in
sensitivity to a chameleonic sector, could be very significant.

We would like to stress again that the afterglow phenomenon in the
experiments considered here is a smoking gun for a chameleonic
particle. From the viewpoint of optical experiments, a number of
further mechanisms have been proposed that could induce optical
signatures in laboratory experiments, but still evade astrophysical
constraints
\cite{Masso:2006gc,Mohapatra:2006pv,Jain:2006ki,Foot:2007cq}. Distinguishing
between the various scenarios in the case of a positive signal for,
say, ellipticity and dichroism, may not always be possible with the
current laboratory set-ups. But the observation of an afterglow would
strongly point to a chameleon mechanism.

In this sense, it appears worthwhile to reconsider other concepts on
using optical signatures to deduce information about the underlying
particle-physics content, e.g., using strong laser fields
\cite{Heinzl:2006xc,DiPiazza:2006pr,Marklund:2006my} or astronomical
observations \cite{Dupays:2005xs, tomi1,tomi2, Mirizzi:2007hr,DeAngelis:2007yu}, in
the light of a chameleon field. 

Finally, it is interesting to notice that it could, in principle, be
possible to measure the varying chameleon mass $m$ by varying the
vacuum pressure in the experiment and thereby extract information
about the scalar potential $V(\phi)$.  The reconstruction of $V(\phi)$
from this afterglow experiment assumes however that the value of $M$
that one detects from afterglow experiments (i.e, the photon-chameleon
coupling) is the same as the matter-chameleon coupling. This, of
course, need not be the case; although we might expect them to be of
the same magnitude. By contrast, the reconstruction of $V(\phi)$ from
the Casimir force tests \cite{chamcas} does not run into this problem,
since these experiments would effectively measure $V(\phi)$ directly
and provide only weak constraints on $M$. If chameleon fields were to
be detected, one could, by comparing the reconstructions of $V(\phi)$
from afterglow and Casimir experiments, actually measure not only the
chameleon-photon coupling but also the chameleon-matter
coupling.

In resume, our afterglow estimates clearly indicate that the
new-physics parameter range accessible even with current technology is
substantial. Afterglow searches therefore represent a powerful tool
to probe physics halfway up to the Planck scale.

\acknowledgments

We thank P. Brax and C. van de Bruck for useful discussions. 
HG acknowledges support by the DFG under contract
Gi 328/1-4 (Emmy-Noether program). 
DFM is supported by the Alexander von Humboldt Foundation. DJS is
supported by STFC. 

\appendix

\section{Interference effects}

In this appendix, we discuss the possible use of interference effects
for the afterglow phenomenon. Whereas the short-pulse experiments
considered in the main text provide for a clean signal, the occurrence of
interference depends more strongly on the details and the precision of
the experiment. 

Let us consider a long pulse of duration $T\gg 2(L+d)$ with frequency
$\omega$ and a Gau\ss ian envelope,
\begin{equation}
a_{\text{in}}(t,z)=a_0 \, e^{-i \omega t+i kz}\, e^{-\frac{1}{2}
  \frac{(t-z)^2}{T^2}}. \label{A1}
\end{equation}
In this case, the afterglow amplitude in \Eqref{Aout1} or \Eqref{Aout2}
for a given time $t$ at position $z$ (say, at a time $t>T$ after the
initial pulse has passed the detector at $z\simeq L+d$) receives
contributions from many terms in the $n$ sum; typically, the 
Gau\ss ian profile picks up a wide range of large $n$ values. 

Extending the summation range of $n$ from $-\infty$ to $\infty$
(instead of $1$ to $\infty$) introduces only an exponentially small
error owing to the Gau\ss ian envelope, but allows to perform a
Poisson resummation of the $n$ sum; for instance, the term in the
second line of \Eqref{Aout2} yields ($\bar\omega=\omega$)
\begin{eqnarray}
&&\sum_{n=1}^{\infty} e^{-i\frac{m^2
      ((2n+1)L+2n d)}{{\omega}}} a_{\text{in}}^{\prime
    \prime}\left(t-(2n+1)(L+d)  - \frac{m^2
    ((2n+1)L+2nd)}{2{\omega}^2}\right) \nonumber\\
&&\to -\sqrt{\frac{\pi}{2}} a_0\, \omega^2 e^{- i\omega t+ ikz + i
      \frac{m^2}{2\omega} d + i \omega (L+d) f_-} 
  \frac{T}{(L+d)f_+}  e^{i \omega \frac{f_-}{f_+} (t-z+\frac{1}{2}
      \frac{m^2}{\omega^2} d -(L+d) f_+) } \nonumber\\
&&\quad \times \sum_{m= -\infty}^\infty e^{i \frac{\pi m}{(L+d)f_+}
      ( t-z+ \frac{1}{2} \frac{m^2}{\omega^2} d -(L+d) f_+)}
  e^{-\frac{1}{2} \frac{T^2}{4(L+d)^2 f_+^2} [ 2\omega (L+d) f_- -2\pi
      m]^2},
\label{A2}
\end{eqnarray}
where $f_\pm= 1\pm \frac{1}{2} \frac{m^2}{\omega^2}$, and we have
dropped terms of order $t/(T^2\omega)$ and $1/(T\omega)^2$. The Gau\ss
ian factor in the last term causes a strong exponential suppression,
since $T\gg 2(L+d)$ and $(L+d)\omega\gg 1$ by assumption, and typically
$f_\pm=\mathcal O(1)$. Therefore, this factor essentially picks out
one term in the $m$ sum that comes closest to the resonance condition
\begin{equation}
m=\frac{\omega(L+d) f_-}{\pi} .\label{Ares}
\end{equation}
Let us denote the integer that is closest to this resonance condition
with $m_{\text{res}}$. Then the afterglow amplitude reads
\begin{equation}
a_{\text{glow}}(t,z)= - a_0  \sqrt{\frac{\pi}{2}}\, \frac{4 B^2
    \omega^2}{M^2 m^4}  \frac{T}{(L+d)f_+} \sin^2 \left[ \frac{m^2
    L}{4\omega^2} \left( \omega \frac{f_-}{f_+} + \pi m_{\text{res}}
    \right) \right] e^{-\frac{1}{2} \frac{T^2}{4(L+d)^2 f_+^2} \left(
    2\omega (L+d)f_- -2\pi m_{\text{res}} \right)^2} \, e^{i\varphi(t,z)},
\label{A5}
\end{equation}
where $\varphi(t,z)$ summarizes all phases of the
amplitude. Whether or not the resonance condition can be met is a
delicate experimental issue. The above formula suggests that the
experimental parameters require an extraordinary fine-tuning, since
the width of the resonance is extremely small. However, in a real
experiment, systematic uncertainties may invalidate the
idealized scenario from the very beginning; for instance, the laser beam
has a finite cross section, and the length of the vacuum chamber $L+d$
may vary across this cross section a little bit. Variations on the
order of the laser wave length lead to uncertainties of order 1 on the
right-hand side of \Eqref{Ares}. Therefore, the resonance condition
might be satisfied for some part of the beam only. 

For a simplified estimate, let us assume that the beam cross section
is a circular disc. The center of the disc is assumed to satisfy the
resonance condition exactly, but the length $L+d$ varies linearly from
the center to the edge of the disc by a bit less than half a laser
wavelength $\lesssim \lambda/2$. Averaging over the Gau\ss ian
resonance factor in \Eqref{A5} then corresponds to integrating
radially over the resonance peak. This procedure leads to the
replacement 
\begin{equation}
 e^{-\frac{1}{2} \frac{T^2}{4(L+d)^2 f_+^2} \left(
    2\omega (L+d)f_- -2\pi m_{\text{res}} \right)^2}
 \to \frac{8}{(T\omega)^2} \left( \frac{f_+}{f_-} \right)^2
    \frac{(L+d)^2}{\lambda^2} = \frac{2}{\pi^2} \frac{(L+d)^2}{T^2}
    \left( \frac{f_+}{f_-} \right)^2 .
\label{A10}
\end{equation}
(More resonance points in the beam cross section would give a sum of
similar terms on the right-hand side of \Eqref{A10}.)  In this case,
we can read off the probability amplitude for a photon in the Gau\ss
ian pulse to reappear in the afterglow at time $t>T\gg (L+d)$ averaged
over the resonance, 
\begin{eqnarray}
\mathcal{P}_{\text{average}} &\simeq& \frac{8}{\pi^2}
\sqrt{\frac{\pi}{2}} 
\frac{B^2 \omega^2}{M^2 m^4} \frac{(L+d)}{T} \frac{f_+}{f_-^2} \sin^2 \left[ \frac{m^2
    L}{4\omega^2} \left( \omega \frac{f_-}{f_+} + \pi m_{\text{res}}
    \right) \right] \nonumber\\ 
 &\simeq& \frac{4}{\pi^2}
\sqrt{\frac{\pi}{2}} 
  \frac{B^2 \omega^2}{M^2 m^4}
 \frac{(L+d)}{T} \frac{1+ \frac{1}{2} \frac{ 
  m^2}{\omega^2}}{(1- \frac{1}{2} \frac{ 
  m^2}{\omega^2})^2},   \label{Pres}
\end{eqnarray}
where we have approximated the $\sin^2$ by its phase average $1/2$,
since the phase is large and may also vary spatially or over the
measurement period. 

Let us compare the resulting probability for a long pulse $T=10\;$s with
that of a short-pulse set-up for $B=5$T, $L\simeq L+d=6$m,
$\omega=2\pi/(532$nm) in the region of small masses $m\lesssim
0.1$meV,
\begin{eqnarray}
\mathcal{P}_{\text{short pulse}}^2 &=& \left[\frac{4B^2 \omega^2}{M^2 m^4} \sin^2\left(\frac{m^2
  L}{4\omega}\right)\right]^2 \simeq \frac{B^4 L^4}{16 M^4} \simeq 3.1\times
  10^{-21} \left(\frac{10^6 \text{GeV}}{M}\right)^4, \nonumber\\  
\mathcal{P}_{\text{average}}^2 &=& \left[\frac{4}{\pi^2}\sqrt{
  \frac{\pi}{2} }
 \frac{B^2 \omega^2}{M^2  m^4}  \frac{(L+d)}{T} \frac{1+ \frac{1}{2} \frac{ 
  m^2}{\omega^2}}{(1- \frac{1}{2} \frac{ 
  m^2}{\omega^2})^2}  \right]^2 \simeq
2.8\times 10^{-33}  \left(\frac{10^6 \text{GeV}}{M}\right)^4
  \left(\frac{0.1 \text{meV}}{m}\right)^8 .
\label{A8}
\end{eqnarray}
Here it is important to stress that the calculation has been performed
in the limit $\theta \simeq \omega B/(m^2 M)\ll 1$, corresponding to 
\begin{equation}
1\gg \theta \simeq 2.2 \times 10^{-4}  \left(\frac{10^6 \text{GeV}}{M}\right)
  \left(\frac{0.1 \text{meV}}{m}\right)^2 . \label{vallim}
\end{equation}
In other words, the validity range of Eqs.~\eqref{Pres} and \eqref{A8}
does not extend to arbitrarily small masses.  Still, for small
chameleon masses $m$ and larger values of $M$ (such that
\Eqref{vallim} holds), the averaged resonance probability can exceed
that of the short-pulse case in the present example. 

Using, for instance, a 10s long pulse from a 100Watt laser, the number
of photons in the afterglow will be
\begin{equation}
n_{\text{glow}} \simeq 2\times 10^{-4}  \left(\frac{10^6 \text{GeV}}{M}\right)^4
  \left(\frac{0.1 \text{meV}}{m}\right)^8  \left(
  \frac{t_{\text{expt}}}{1 \text{s}} \right),
\end{equation}
where $t_{\text{expt}}$ denotes the measurement time of the
afterglow. We conclude that such an experiment can become sensitive to
the coupling scales of order $M\sim 10^{10}$GeV for chameleon masses
in the sub $\mu$eV range.  We stress again that the precise
sensitivity limits strongly depend on the experimental details and the
feasibility to exploit a chameleon resonance in the vacuum cavity. If
such a resonance can be created in an experiment the interference
measurement can provide for a handle on a measurement of the chameleon mass,
since the dependence on the latter is rather strong. 

If a fully controlled resonance could be built up in an experiment,
the suppression factor on the right-hand side of \Eqref{A10} arising
from averaging would be replaced by 1; this would correspond to an
enhancement factor of $10^{18}$ for the probability amplitude in the
example given above. From an experimental viewpoint, controlling the
resonance a priori seems, of course, rather difficult. From the
non-observation of an afterglow in a long-pulse experiment, it would
be difficult to conclude whether this points to rather strong bounds
on $M$ or to the fact that the resonance condition is not sufficiently
met. We therefore recommend short-pulse experiments as a much cleaner
and well controllable set-up.

\end{document}